\documentclass[aps,nofootinbib,preprint]{revtex4}
\usepackage{amsmath,amssymb}
\usepackage{graphicx}

\def\bi{\boldsymbol i}
\def\bq{\boldsymbol q}

\begin{document}

\title{
A Class of Five-Dimensional Multi-Charged Tchrakian Monopoles
}

\date{\today}

\author{Hironobu Kihara\thanks{e-mail:hkihara@th.phys.titech.ac.jp}}
\affiliation{Tokyo Institute of Technology}

\begin{abstract}
By considering singular gauge transformation of 
the five-dimensional Hedge-Hog monopole, 
we obtain a class of multi-charged Tchrakian monopoles. 
The charge is classified by the fourth homotopy group 
of SU(4)/U(1)$\times$SU(2)$\times$SU(2). 
\end{abstract}

\maketitle

~~Magnetic monopoles and its generalization have been 
considered for long time \cite{Dirac:1948um}. 
A finite energy monopole in five-dimensional space 
is considered by Tchrakian 
as a generalization of the 't Hooft-Polyakov monopole, 
in spaces whose dimensions are greater than four 
\cite{Tchrakian:1978sf}. 
We study a class of multi-charged monopoles in 
five-dimensional space in this brief article. 
The mapping class degrees obtained by scalar field configurations 
in the asymptotic region are identified with charges of those configurations 
\cite{Arafune:1974uy}. 

Let us consider monopoles in SU(4) gauge theory 
in five-dimensional space. 
The pseudo-energy is defined as 
\begin{eqnarray}
E &=& \frac{1}{8} \int  {\rm Tr} 
\left\{ ( F \wedge F ) \wedge *  ( F \wedge F ) 
- D \phi \wedge * D \phi  \right\}  ~.
\end{eqnarray}
Here $F$ is a su(4)-valued gauge field strength and 
$\phi$ is a su(4)-valued scalar field. 
Generators of SU(4) are represented as anti-Hermitian. 
$*$ is denoted the Hodge dual operator in five-dimensional
 Euclidean space.  
Suppose that the value of the scalar field in asymptotic 
region is stabilized to a constant $H_0$ by scalar potential 
and we consider Prasad-Sommerfield limit. 
We omit the scalar potential. 
We study solutions of the following Bogomol'nyi equation, 
\begin{eqnarray}
F \wedge F &=& \pm * \bi  D \phi ~.
\label{eqn:bogo}
\end{eqnarray}

For a while we list our notations.
Points in the space are parametrized by five real 
numbers $( x_1 , x_2 , x_3 , x_4 , x_5 )$. 
Let us define the Pauli matrices, 
\begin{eqnarray}
\sigma_1 &=& 
\begin{pmatrix} 0 & 1 \\ 1 & 0 \end{pmatrix} ~,\cr
\sigma_2 &=& 
\begin{pmatrix} 0 & - \bi \\ \bi & 0 \end{pmatrix} ~,\cr
\sigma_3 &=& 
\begin{pmatrix} 1 & 0 \\ 0 & -1 \end{pmatrix} ~. \nonumber
\end{eqnarray}
These matrices satisfy the following multiplication relations, 
$\sigma_{a} \sigma_{b} = \delta_{ab} {\bf 1}_2 
+  \bi \epsilon_{abc} \sigma_c $ and commutation relation
$[\sigma_a , \sigma_b] = 2 \bi \epsilon_{abc} \sigma_c ~,$
where $a,b,c= 1,2,3$. 
By using these matrices the generators of the 
five-dimensional Clifford algebra are defined as 
\begin{eqnarray}
\gamma_1 &=& \sigma_1 \otimes  \sigma_2 ~,\cr
\gamma_2 &=& \sigma_2 \otimes  \sigma_2~,\cr
\gamma_3 &=& \sigma_3 \otimes  \sigma_2 ~,\cr 
\gamma_4 &=& {\bf 1}_2 \otimes \sigma_1 ~,\cr
\gamma_5 &=& {\bf 1}_2 \otimes \sigma_3 ~,\nonumber
\end{eqnarray}
These matrices satisfy the anticommutation relation, 
$\{ \gamma_M , \gamma_N \} = 2 \delta_{MN}$, $(M,N = 1,2,3,4,5)$. 
We have one more relation, 
$\gamma_1 \gamma_2 \gamma_3 \gamma_4 \gamma_5 = {\bf 1}_4 $. 
Matrices $\bi \gamma_a , \gamma_{ab}$ are closed under commutation. 
They form the Lie algebra su(4). 
When the scalar field has nonzero vacuum expectation value, 
the gauge transformation is broken. 
Suppose that $\phi = \bi  H_0 \gamma_5$ at infinity, 
where $H_0$ is a positive real constant. 
generators $\bi \gamma_5, \gamma_{ij}$ commute with $\phi$, 
while generators $\bi \gamma_i, \gamma_{i5}$ anti-commute with $\phi$. 
Here $i,j=1,2,3,4$. 
Generators $\bi \gamma_5, \gamma_{ij}$ span the unbroken 
symmetry U(1)$\times$SU(2)$\times$SU(2).

The radial coordinate of the five dimensional space 
is defined as $r^2 = x_M^2 $. 
Let us use the following spherical coordinate system, 
\begin{eqnarray}
x_1 &=& r \sin \theta y_1 ~,\cr
x_2 &=& r \sin \theta y_2 ~,\cr
x_3 &=& r \sin \theta y_3 \cr
x_4 &=& r \sin \theta y_4 ~,\cr
x_5 &=& r \cos \theta  ~.\nonumber
\end{eqnarray}
Here $y_i,~~ (i=1,2,3,4)$ is a point on $S^3$ with unit radius, 
$(y_1)^2 +  (y_2)^2+  (y_3)^2+  (y_4)^2 =1$. 

The following SU(2) matrix plays 
the crucial role in this article, 
\begin{eqnarray}
w &:=& y_4 {\bf 1}_2 + y_1 \bi \sigma_1 + y_2 \bi \sigma_2 + y_3 \bi \sigma_3  ~.
\end{eqnarray}
The quantity $w$ is singular at $\rho =0$ or $\theta = 0 , \pi$. 
The Hedge-Hog monopole is written in terms of 
\begin{eqnarray}
e_{(1)} :=  \frac{1}{r} x_M \gamma_M 
&= &
\begin{pmatrix} 
\cos \theta {\bf 1}_2 & \sin \theta {w^{\dag}}  \\
\sin \theta w & - \cos \theta {\bf 1}_2
\end{pmatrix} ~.
\end{eqnarray}
The matrix $e_{(1)}$ is defined everywhere except for $x_M=0$. 
$e_{(1)}$ satisfies the following properties, 
\begin{eqnarray}
e_{(1)}^2 &=& {\bf 1}_4~,\cr
* d e_{(1)}^4 &=& -\frac{4!}{r^4} edr~,\cr
* dr \wedge d e_{(1)}^3 &=& - \frac{3!}{r^2} e_{(1)}  d e_{(1)} ~.
\end{eqnarray}
The matrix $e_{(1)}$ is transformed from $\gamma_5$ with a unitary matrix ${\cal U}_{(1)}$,
\begin{eqnarray}
e_{(1)} &=& {\cal U}_{(1)} \gamma_5 {\cal U}_{(1)}^{-1} ~,\cr
{\cal U}_{(1)}&:=& V R V^{-1} ~,\cr
R &:=& \begin{pmatrix}
\cos (\theta/2) {\bf 1}_2 & -\sin (\theta/2)  {\bf 1}_2\\
 \sin (\theta/2) {\bf 1}_2 & \cos (\theta/2)  {\bf 1}_2
\end{pmatrix}~,\cr
V &:=& \begin{pmatrix}
1 & 0 \\ 
0 & w
\end{pmatrix} ~.
\end{eqnarray}
The matrix $V$ is singular at $\theta = 0 , \pi$. 
However at $\theta=0$, $R={\bf 1}_4$ and ${\cal U}_{(1)}={\bf 1}_4$. 
Hence, ${\cal U}_{(1)}$ is regular at $\theta=0$. 
At $\theta =\pi$ the matrix ${\cal U}_{(1)}$ is singular. 
The configuration of the Hedge-Hog monopole is given as
\begin{eqnarray}
\phi_{(1)} &=& \bi H_0 U(r) e_{(1)}~,\cr
A_{(1)} &=& \frac{1-K(r)}{2 \bq} e_{(1)} d e_{(1)} ~.
\end{eqnarray}
Boundary conditions, $U(0)=0$ and $K(0)=1$, make these configuration 
well-defined at the origin. 
This configuration satisfy the Bogomol'nyi equation (Eq.(\ref{eqn:bogo})), 
\begin{eqnarray}
F_{(1)} \wedge  F_{(1)} &=& -  * \bi D_{(1)} \phi_{(1)} ~.
\end{eqnarray}
Here $F_{(1)}= d A_{(1)} + \bq A_{(1)}^2$. 

As an extension of $e_{(1)}$, let us use 
\begin{eqnarray}
e_{(n)} &:=& \begin{pmatrix} 
\cos \theta {\bf 1}_2 & \sin \theta {w^{\dag}}^n  \\
\sin \theta w^n & - \cos \theta {\bf 1}_2
\end{pmatrix} ~.
\end{eqnarray}
The square of  $e_{(n)}^2$ is  ${\bf 1}_4$. 
The matrix $e_{(n)}$ is obtained by gauge transformation of $\gamma_5$, too. 
\begin{eqnarray}
e_{(n)} &=& {\cal U}_{(n)} \gamma_5 {\cal U}_{(n)}^{-1} ~,\cr
{\cal U}_{(n)}&:=& V^n R V^{-n}~.
\end{eqnarray}
At $\theta =0$, $R={\bf 1}_4$ and ${\cal U}_{(n)} = {\bf 1}_4$, 
while at $\theta=\pi$ the matrix ${\cal U}_{(n)}$ is singular. 

Let us consider the following configuration,  
\begin{eqnarray}
\phi_{(n)} &=& \bi H_0 U(r) e_{(n)}~,\cr 
A_{(n)} &=& 
\left( {\cal U}_{(n)} {\cal U}_{(1)}^{-1} \right)
A_{(1)} 
\left( {\cal U}_{(n)} {\cal U}_{(1)}^{-1} \right)^{-1}
+ \frac{1}{\bq} \left( {\cal U}_{(n)} {\cal U}_{(1)}^{-1} \right)
    d \left( {\cal U}_{(n)} {\cal U}_{(1)}^{-1} \right)^{-1} ~.
\end{eqnarray}
 The corresponding field strength is given as,
\begin{eqnarray}
F_{(n)} &=&  \left( {\cal U}_{(n)} {\cal U}_{(1)}^{-1} \right)
 F_{(1)} \left( {\cal U}_{(n)} {\cal U}_{(1)}^{-1} \right)^{-1} ~. 
\end{eqnarray}
The configuration satisfies the Bogomol'nyi equation, too, 
\begin{eqnarray}
* F_{(n)} \wedge F_{(n)} &=& \left( {\cal U}_{(n)} {\cal U}_{(1)}^{-1} \right)
 ( * F_{(1)} \wedge F_{(1)} ) \left( {\cal U}_{(n)} {\cal U}_{(1)}^{-1} \right)^{-1}\cr
&=& -\bi \left( {\cal U}_{(n)} {\cal U}_{(1)}^{-1} \right)
 D_{(1)} \phi_{(1)} \left( {\cal U}_{(n)} {\cal U}_{(1)}^{-1} \right)^{-1} \cr
&=& - \bi D_{(n)} \phi_{(n)} ~. 
\end{eqnarray}
Here $F_{(n)}= d A_{(n)} + \bq A_{(n)}^2$. 

Let us compute the topological charge of this configuration, 
\begin{eqnarray}
Q_n &:=& - \frac{\bi}{{\rm vol}(S^4)} \int_{S^4}{\rm Tr} \hat{\phi}_{(n)} d\hat{\phi}_{(n)}^4 
 = 4 \cdot 4! n~, \cr
| \phi | & :=& \sqrt{ \frac{1}{4} {\rm Tr} \phi^{\dag} \phi} \cr
\hat{\phi} &:=& \frac{\phi}{| \phi |} ~.
\end{eqnarray}
This quantity is classified by the fourth homotopy group 
$\pi_4(SU(4)/U(1) \times SU(2)\times SU(2)) \simeq {\rm Ker}\{ \pi_3(U(1) \times SU(2) \times SU(2)) \rightarrow \pi_3(SU(4))  \} \simeq \mathbb{Z}$.
Hence the configuration is not topologically equivalent to the Hedge-Hog configuration. 
We have constructed a simple class of the multi-charged Tchrakian monopoles. 

{\bf Acknowledgment}\\
The author would like to thank to Muneto Nitta, Shinji Shimasaki and Shin Sasaki for their comments. 


\end{document}